\documentclass[nonacm,sigplan]{acmart}
\renewcommand\footnotetextcopyrightpermission[1]{} 
\usepackage[utf8]{inputenc}
\makeatletter%
\@ifclassloaded{acmart}{
  \makeatletter
  \def\mdseries@tt{m}
  \makeatother
}{}
\makeatother

\usepackage{amsmath,mathtools}
\usepackage{bbm}
\usepackage{listings} 
\usepackage{float}
\usepackage{hyperref}
\usepackage{microtype}
\usepackage{tikz}
\usetikzlibrary{decorations.text,calc,shapes,arrows,arrows.meta, positioning,shapes.misc,decorations.markings,decorations.markings,decorations.pathreplacing,matrix}

\usepackage{booktabs}
\usepackage{multirow}
\usepackage{adjustbox}
\usepackage{verbatim}
\usepackage[T1]{fontenc}
\usepackage[title]{appendix}
\usepackage{adjustbox} 
\usepackage{graphicx}
\usepackage{caption}
\usepackage{subcaption}
\usepackage{algpseudocode}
\usepackage{algorithm}
\usepackage{siunitx}
\usepackage{nicefrac}
\usepackage{enumitem}
\usepackage[colorinlistoftodos]{todonotes}
\newcommand{\textcite}[1]{\citet{#1}}

\usepackage[autostyle, english=american]{csquotes}
\MakeOuterQuote{"}


\newcolumntype{t}{>{\ttfamily}l}
\newcolumntype{T}{>{\ttfamily}c}

\newcolumntype{^}{>{\currentrowstyle}}

\everypar{\looseness=-1} 
\hypersetup{draft}
\setcopyright{none}
\begin{document}


\title[]{A Feature Set of Small Size for the PDF Malware Detection}

  \author{Ran Liu}
  \affiliation{%
    \institution{Univ. of Maryland, Baltimore County}
  }
  \email{rliu2@umbc.edu}

  \author{Charles Nicholas}
  \affiliation{%
    \institution{Univ. of Maryland, Baltimore County}
  }
  \email{nicholas@umbc.edu}



\begin{abstract}
Machine learning (ML)-based malware detection systems are becoming increasingly important as malware threats increase and get more sophisticated. PDF files are often used as vectors for phishing attacks because they are widely regarded as trustworthy data resources, and are accessible across different platforms. Therefore, researchers have developed many different PDF malware detection methods. Performance in detecting PDF malware is greatly influenced by feature selection. In this research, we propose a small features set that don't require too much domain knowledge of the PDF file. We evaluate proposed features with six different machine learning models. We report the best accuracy of 99.75\% when using Random Forest model. Our proposed feature set, which consists of just 12 features, is one of the most conciseness in the field of PDF malware detection. Despite its modest size, we obtain comparable results to state-of-the-art that employ a much larger set of features.
\end{abstract}


\maketitle
\pagestyle{plain}

\section{Introduction}
The flexibility and portability of PDF files makes them a popular target for malware attacks. Over time, different approaches have been proposed to detect PDF malware. Machine learning and neural network based models have particularly shown promise in these detection tasks. However, the performance of the model relies on the quality of the feature set chosen\cite{Liu2023CanDetection}.  Features used in malware detection are grouped into two categories: dynamic and static. Dynamic features are obtained from monitoring program execution, such as APIs called, instructions executed, or IP addresses accessed. Conversely, static features are obtained through static analysis. Both categories have some limitations. Dynamic features need to be executed in a sandbox environment, where some sophisticated malware can detect the sandbox environment and consequently alter their behaviors. Static features, on the other hand, can be obfuscated by attackers using evasion techniques, making the detection challenging. This raises concerns that some commonly used features have been thoroughly investigated by attackers. If attackers have exploited these features to perform a successful evasive attack, PDF malware detection systems built on the same or similar features set might become vulnerable. This highlights the importance of the usage of PDF-specific features, which may reduce the attack surface. Earlier research, including PDFRate, have employed some PDF-specific features such as the number and the occurrence of a specific PDF objects for model training, which obtained promising accuracy in PDF malware detection\cite{ParkourDump.}\cite{Maiorca2015AFiles}. Nevertheless, most of these features requires a large amount of domain knowledge to extract. Moreover, their feature sets are large and complex, which may potentially lead to over-fitting. Consequently, it's desirable to have a simple and small PDF-specific features set that may achieve detection accuracy comparable to more complex features.

In this paper, we limit the scope of PDF-specific features to those that are unique to PDF files, hence excluding most dynamic features such as system call sequences, API call sequences, and some static features such as binary code.  Furthermore, we exclude features that need extensive domain knowledge for PDF files, which means that most keyword-based features, including JavaScript code in PDF files, are not used in our research.

PDF files can be viewed as a set of interconnected objects. Some work, such as Hidost's, extracts the tree structure of the PDF and uses the binary counts for these paths as features\cite{Srndic2016Hidost:Files}.  Our previous work showed that such tree structures contain sequential relationships and can be used to train the Time Series Model for PDF malware detection.  In this paper, we propose a novel set of graph features to accurately detect PDF malware. We investigated multiple types of graph tree features by parsing a PDF file into tree representative features. For specific feature types, our research demonstrated statistical differences between benign and malicious PDFs. Using the proposed feature set to train a machine learning model, we show empirically that the model can successfully detect 99.75\% of PDF malware samples with only 12 features. The primary contributions of our study are the introduction of one of the smallest PDF specific feature sets. We have conducted a thorough investigation and performance analysis of the ML - models based on these proposed features. Furthermore, we benchmarked our results against state-of-the-arts, indicating that our feature set is promising.

\subsection{PDF file Structure}
\begin{figure}
    \centering
    \includegraphics[scale=0.2]{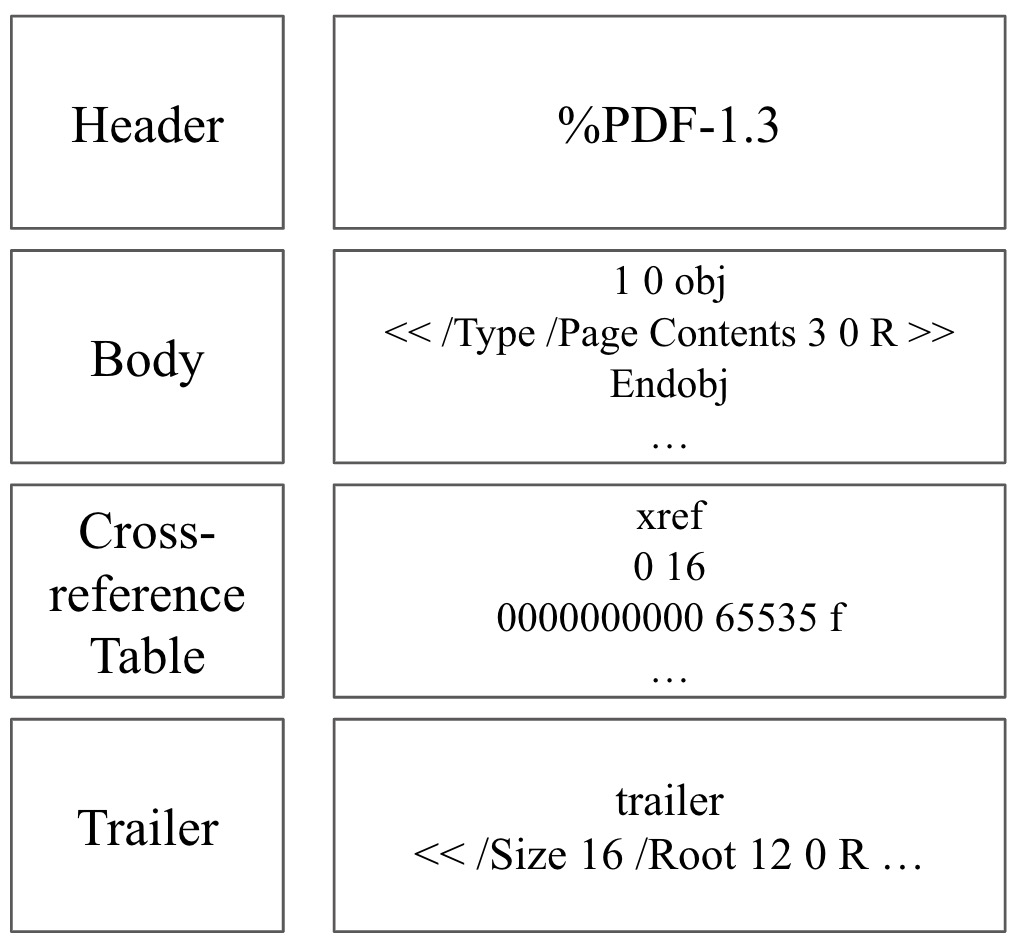}
    \caption{PDF Structure}
    \label{fig:PDFStructure}
\end{figure}

A PDF file is structured using interconnected modules known as objects, which are made up of four parts: a header, a body, a cross-reference table, and a trailer, as shown in \autoref{fig:PDFStructure}. 
\begin{itemize}
    \item Header: The header contains information about the PDF version and is marked with the '\%' symbol.
    \item Body: The body, as the primary section of the PDF file, consists of objects that define all the operations performed by the file. These objects, which include both indirect and direct objects, characterize the functionality of each object using keywords marked with '/'. For example, '/Length' and '/Filter' are such keywords. Indirect objects, which start with a numeric identifier like '4 0 R', contain information in a directory and can be referenced by other objects. For example, an object starting with '1 0 R' can be referred to by other objects using '1', the sequence number. This structure allows for the interconnection of objects. The generation number, usually set to 0, is represented by the second digit, although it can be other number in some cases. Objects are usually end with the 'endobj' marker. A specific type of object, known as a stream, starts with the keyword 'stream' and ends with 'endstream' and 'endobj'. The content of the stream object, which includes elements like images and texts, is encoded using filters.
    \item Cross-reference table: This table contains the location references for each object. A PDF parser uses this table to find the object reference in the memory for parsing. The cross-reference table, marked by 'Xref' followed by numbers, indicates the total number of objects in the references with its last number. For instance, '0 16' indicates a total of 16 objects in the cross-reference table.
    \item Trailer: trailer contains information about the file such as the number of objects using keyword '/Size'. It also contains a reference to the root object using keyword /Root and metadata using keyword /Info.
    The file structure organizes the logical access order for the PDF file. When a PDF reader application accesses a PDF file, it first locates the trailer to find the root object. Then the parser uses a cross-reference table to parse each indirect object to decompress all data. In this way, the content of the PDF is made visible to the user. When a modification happens in the PDF file, like inserting a page into the PDF,  a new body, trailer, and cross-reference table will be appended to the original file accordingly, and a new version number will be generated. However, because the cross-reference table sets a strict boundary for each object, removing objects from the previous version can cause errors. Thus, an attacker is more likely to add features instead of removing features.
\end{itemize}
Each object is labeled with a number, allowing it to be referenced by other objects. Object information is stored in the cross-reference table, while the trailer indicates the root object's number and the cross-reference table's location. By querying the cross-reference table, catalog objects can be found. The catalog object serves as the entire document's root object, containing the PDF document's outline and the page group object's reference.



\subsection{Related Work}

In the field of machine learning-based PDF malware detection, two primary types of features are commonly used - static and dynamic features. The dynamic features are obtained by running the PDF in a controlled environment, which allows for the collection of PDF running behaviors such as sequences of system calls and API calls\cite{RieckAutomaticLearning}\cite{Liu2023IMCDCF:Models}. However, most dynamic features are typically not distinctive to PDF files, and the building of a robust sandbox environment increases the complexity of the detection process. Despite the promising results in malware detection tasks using dynamic features, our work primarily focuses on the use of static features. Previously used dynamic and static features require significant amounts of domain knowledge for feature extraction. In contrast, our goal is to employ features that demand minimal domain knowledge and are unique to PDF files.

There are three categories of static features. The first type of static features are obtained from a keyword-based analysis, which involves searching for predefined keywords such as '/Javascript', '/OpenAction', '/GoTo', '/URI', and '/RichMedia'.  These keywords are often associated with malicious code injection. Features can include the number of keywords or simply their presence. The malicious payloads are usually inserted into objects associated with such keywords, making them useful features for PDF malware detection. The second type of static features are obtained through a tree structure-based analysis. This method constructs the object tree representation to capture the connections between items. The tree structure can provide insights into the hierarchical connections between the objects of the PDF file, which may reveal malware-related behaviors. Lastly, static features can be obtained using code-based analysis, which focuses on malicious strings and functions in Javascript code. As PDF malware often manipulates Javascript code to execute malicious activities, the presence of specific code strings or functions may indicate the presence of malware. PDF malware detectors may employ one or more of the features described above. One such example is Hidost\cite{Srndic2016Hidost:Files}, a system that uses the Poppler PDF parser\cite{FreeDesktop.org.2018Poppler} to extract tree structural paths of objects in a PDF file, which are then used as features in the classification process. Hidost is implemented with two different models: Support Vector Machine (SVM) and Random Forest (RF)\cite{vsrndic2013detection}\cite{Srndic2016Hidost:Files}. SVM is a supervised learning model that creates an optimal hyperplane to separate different labels. RF is a meta-estimator that integrates different decision trees to improve classification accuracy. The researchers trained their model using a dataset of 10,000 randomly selected files, maintaining a malicious-to-benign ratio of 1:1. The complete PDF dataset comprised 407,037 benign and 32,567 malicious files. The Hidost system has 99.8\% accuracy and less than 0.06\% false positive rate for both models.

The PDFrate classifier is implemented using an RF algorithm with 99\% accuracy and 0.2\% false positive rate over the Contagio malware dataset\cite{ParkourDump.}. PDFrate uses the metadata, which includes the names of the files' authors, the size of the file, its location, and the amount of certain keywords, and the content of the PDF files as features. The authors manually define the feature set, which has 202 features in all, including counts for different keywords and specific fields in the PDF. Examples include the number of characters in the author field, the quantity of "endobj" keywords, the total number of pixels in all the photos, the quantity of JavaScript markers, etc.  The Mimicus implementation of PDFrate, claiming to get a close approximation, only makes use of 135 of these features. The two versions of PDFrate, PDFrate-v1 and PDFrate-v2, each use a different machine learning model\cite{Yerima2022MaliciousSet}\cite{Smutz2016WhenDetectors}. The classifiers in PDFrate-v2 use mutual agreement to implement an ensemble technique. The term "uncertain" is introduced into the classifier voting, where rates of 25–50\% are regarded as benign uncertainty and rates of 50–75\% as malicious uncertainty. 

PjScan is a tool that concentrates on examining JavaScript code\cite{Laskov2011StaticDocuments}. It uses Poppler\cite{FreeDesktop.org.2018Poppler} as a parser to extract tokens from JavaScript code, and a one-class SVM as a classifier. PjScan achieved 85\% detection accuracy. Malware Slayer is available in two variants\cite{Maiorca2012ADetection}\cite{Maiorca2015AFiles}. The original Slayer extracts keyword features from PDF files using a pattern recognition method and labels samples using a random forest algorithm. Slayer NEO uses the PeePDF\cite{Tonn2013PhoneyPDF} and Origami\cite{SogetiESECLab2015Origami} parsers to extract structural data as features and the AdaBoost algorithm for classification.

Maryam et al. proposed a PDF malware detection system based on stacking learning\cite{Issakhani2022PDFLearning}. Their approach is based on the idea that combining different classifiers could produce improved accuracy as each classifier operates based on unique data assumptions. Their feature set included ten general features, such as PDF size and title character count, as well as structural features such as the amount of keywords and objects. These extracted features were initially fed into a base layer consisting of SVM, Random Forest, MLP, and AdaBoost. The prediction outputs from this layer were subsequently fed into a meta-layer featuring Logistic Regression, K-Nearest Neighbors, and Decision Trees. Their reported metrics on the hybird dataset including Contagio dataset were impressive: an accuracy of 99.98\%, precision of 99.84\%, recall of 99.89\%, and an F1 score of 99.86\%.

\section{Statistical Analysis of a Feature Set}

We now introduce our proposed feature set. We use the Contagio dataset, which includes 9,000 benign and 10,982 malicious PDF samples, to extract features\cite{ParkourDump.}. This dataset was chosen due to its accessibility and its large number of labeled samples. We used the pdfrw library to extract tree structure paths for each file, and while some samples were corrupted, we were able to successfully extract path objects from  7,396 benign PDF files and 10,814 malicious PDF files\cite{Xu2016AutomaticallyClassifiers}. Our feature selection strategy is to minimize required domain knowledge during feature extraction. Consequently, despite the promising results achieved by keyword-based features in other research, we chose not to use them. We have already noted that a PDF can be represented as a tree structure of objects, prompting our investigation into graph features. Our selection of features was facilitated by comparing mean values, standard deviations with 95\% CI and Quantiles. This led us to select the following features:
\begin{itemize}[noitemsep,topsep=0pt]
    \item Distribution of children per node: the average (avg children), median (median children) and variance of children per node (var children).
    \item Number of leaves in the tree (num leaves).
    \item Number of nodes (num nodes).
    \item The depth of the tree (depth).
    \item Average degree (avg degree).
    \item Degree assortativity coefficient (degree assortativity).\footnote{Degree assortativity refers to the tendency for nodes of high or low degree to be connected to other nodes of high or low degree, respectively.}
    \item The average shortest path length (avg shortest path).
    \item How nodes in a graph tend to cluster together (avg clustering coefficient).
    \item Graph density (density).
\end{itemize}

We applied statistical analysis to investigate the proposed features of benign and malicious PDF files. The key statistical metrics in our investigation were: the 75\% quartile, median 50\%, 25\% quartile, 95\% CI mean and 95\% CI standard deviation. 

\begin{table}[!ht]
\centering
\caption{Quantiles for the Proposed Feature Set for Benign PDFs}
\addtolength{\tabcolsep}{1pt}    
\begin{tabular}{@{}lcrcr@{}}
\toprule
            \multicolumn{2}{c}{Benign PDF}&              Quantiles                          \\ 
           & 75\% & \multicolumn{1}{c}{50\%} & 25\%  \\ \midrule
avg children & \multicolumn{1}{r}{1.9416} & 1.5218 & \multicolumn{1}{r}{1.44}                  \\
avg clustering coefficient    & \multicolumn{1}{r}{0.0640} & 0.0185     & \multicolumn{1}{r}{0.097}               \\
avg degree    & \multicolumn{1}{r}{3} & 3                    & \multicolumn{1}{r}{2}                   \\
avg shortest path & \multicolumn{1}{r}{1.0667} & 0.7118            & \multicolumn{1}{r}{0.5421}                 \\ 
degree assortativity & \multicolumn{1}{r}{-0.3781} & -0.4298            & \multicolumn{1}{r}{-0.5569}
\\
density & \multicolumn{1}{r}{0.0215} & 0.0083            & \multicolumn{1}{r}{0.0068}
\\
depth & \multicolumn{1}{r}{4} & 4            & \multicolumn{1}{r}{4}
\\
median children & \multicolumn{1}{r}{0} & 0            & \multicolumn{1}{r}{0}
\\
num edges & \multicolumn{1}{r}{459} & 309            & \multicolumn{1}{r}{189}
\\
num leaves & \multicolumn{1}{r}{188} & 158            & \multicolumn{1}{r}{71} 
\\
num nodes & \multicolumn{1}{r}{260} & 199            & \multicolumn{1}{r}{135.25} 
\\
var children & \multicolumn{1}{r}{158.2680} & 94.4440            & \multicolumn{1}{r}{51.1347}
\\
\bottomrule
\end{tabular}
\label{tbl:beniQuan}
\end{table}

\begin{table}[!ht]
\centering
\caption{Quantiles for the Proposed Feature Set for Malicious PDFs}
\addtolength{\tabcolsep}{1pt}    
\begin{tabular}{@{}lcrcr@{}}
\toprule
            \multicolumn{2}{c}{Malicious PDF}&              Quantiles                          \\ 
           & 75\% & \multicolumn{1}{c}{50\%} & 25\%  \\ \midrule
avg children & \multicolumn{1}{r}{1.7619} & 1.5758 & \multicolumn{1}{r}{1.5714}                  \\
avg clustering coefficient    & \multicolumn{1}{r}{0.0698} & 0.0413     & \multicolumn{1}{r}{0.0374}               \\
avg degree    & \multicolumn{1}{r}{3} & 3                    & \multicolumn{1}{r}{3}                   \\
avg shortest path & \multicolumn{1}{r}{0.6} & 0.5595            & \multicolumn{1}{r}{0.4542}                 \\ 
degree assortativity & \multicolumn{1}{r}{-0.2821} & -0.3126            & \multicolumn{1}{r}{-0.3405}
\\
density & \multicolumn{1}{r}{0.0881} & 0.0786            & \multicolumn{1}{r}{0.0686}
\\
depth & \multicolumn{1}{r}{4} & 4            & \multicolumn{1}{r}{4}
\\
median children & \multicolumn{1}{r}{1} & 1            & \multicolumn{1}{r}{1}
\\
num edges & \multicolumn{1}{r}{40} & 37            & \multicolumn{1}{r}{33}
\\
num leaves & \multicolumn{1}{r}{10} & 10            & \multicolumn{1}{r}{10} 
\\
num nodes & \multicolumn{1}{r}{27} & 21            & \multicolumn{1}{r}{21} 
\\
var children & \multicolumn{1}{r}{4.1814} & 4.0272            & \multicolumn{1}{r}{3.6734}
\\
\bottomrule
\end{tabular}
\label{tbl:malQuan}
\end{table}

\begin{table}[!ht]
\centering
\caption{95\% CI Mean Comparison for the Proposed Feature Set}
\addtolength{\tabcolsep}{-1.5pt}    
\begin{tabular}{@{}lcrcr@{}}
\toprule
            &\multicolumn{1}{c}{Benign}& Malicious                          \\ 
           & 95\% CI Mean & \multicolumn{1}{c}{95\% CI Mean}  \\ \midrule
avg children & \multicolumn{1}{r}{1.8601 - 1.9044} & 1.6359 - 1.6469                   \\
avg clustering coefficient    & \multicolumn{1}{r}{0.0372 - 0.0392} & 0.0501 -
0.0512               \\
avg degree    & \multicolumn{1}{r}{3.1385 - 3.2299} & 2.8545 - 
2.8762
                    \\
avg shortest path & \multicolumn{1}{r}{0.8079 - 0.8279} & 0.5430 -
0.5498                  \\ 
degree assortativity & \multicolumn{1}{r}{-0.4589 - -0.4541} &  -0.3071 - -0.3033
 
\\
density & \multicolumn{1}{r}{0.0205 - 0.0213} & 0.0764 - 0.0772           
\\
depth & \multicolumn{1}{r}{4.2561 - 4.2831} & 3.9594 - 3.9692          
\\
median children & \multicolumn{1}{r}{0.1506 - 0.1942} & 0.9405 - 
0.9612        
\\
num edges & \multicolumn{1}{r}{441.5762 - 471.6670} & 43.8435 - 
45.6364         
\\
num leaves & \multicolumn{1}{r}{152.1234 - 157.9772} & 12.2298 - 12.7902
          
\\
num nodes & \multicolumn{1}{r}{201.8999 - 209.8423} & 25.4634 - 26.2143        
\\
var children & \multicolumn{1}{r}{114.7908 - 122.0534} & 4.8081 - 5.34606      
\\
\bottomrule
\end{tabular}
\label{tbl:beniStat}
\end{table}

\begin{table}[!ht]
\centering
\caption{95\% CI Standard Deviation Comparison for the Proposed Feature Set}
\addtolength{\tabcolsep}{-1.5pt}    
\begin{tabular}{@{}lcrcr@{}}
\toprule
             &\multicolumn{1}{c}{Benign}& Malicious                          \\ 
           & 95\% CI SD & \multicolumn{1}{c}{95\% CI SD}  \\ \midrule
avg children & \multicolumn{1}{r}{0.9561 - 0.9874} & 0.2881 - 0.2959  \\
avg clustering coefficient    & \multicolumn{1}{r}{0.0427 - 0.0441} & 0.0274 - 0.0282                \\
avg degree    & \multicolumn{1}{r}{1.9738 - 2.0385} & 0.5662 - 0.5815                               \\
avg shortest path & \multicolumn{1}{r}{0.4358 - 0.4500} & 0.1759 - 0.1806
                      \\ 
degree assortativity & \multicolumn{1}{r}{0.1097 - 0.1133} & 0.1005 - 0.1032         
\\
density & \multicolumn{1}{r}{0.0230 - 0.0238} & 0.02047 - 0.0210          
\\
depth & \multicolumn{1}{r}{0.5867 - 0.6059} & 0.2579 - 0.2648        
\\
median children & \multicolumn{1}{r}{0.9462 - 0.9772} & 0.5416 - 0.5563          
\\
num edges & \multicolumn{1}{r}{649.6877 - 670.9714} & 46.9322 - 48.2001          
\\
num leaves & \multicolumn{1}{r}{126.3856 - 130.5260} & 14.6701 - 15.0664         
\\
num nodes & \multicolumn{1}{r}{171.4772 - 177.0948} & 19.6561 - 20.1871       
\\
var children & \multicolumn{1}{r}{156.8088 - 161.9459} & 14.0821 - 14.4626          
\\
\bottomrule
\end{tabular}
\label{tbl:malStat}
\end{table}

\begin{figure}[!ht]
    \centering
    \includegraphics[width = 3.3in]{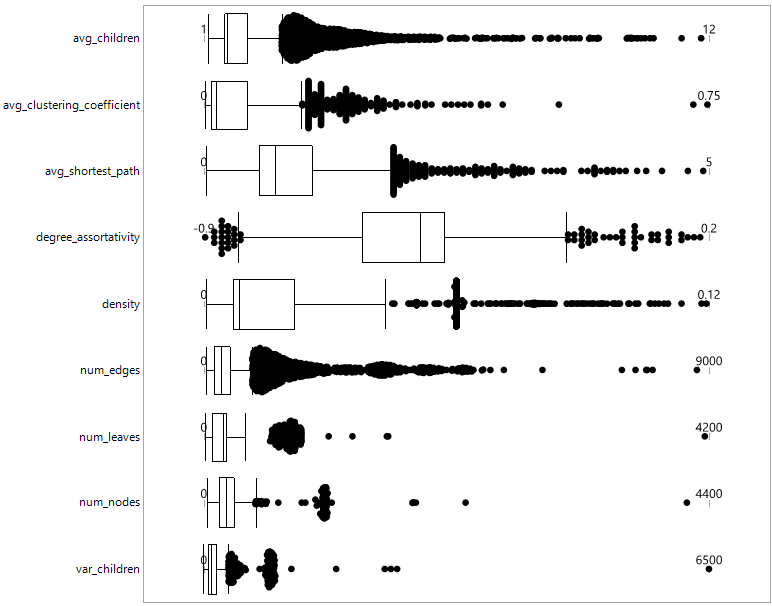}
    \caption{Box plot of the benign PDFs. Features are on the y-axis.  }
    \label{fig:BenPlot}
\end{figure}

\begin{figure}[!ht]
    \centering
    \includegraphics[width = 3.3in]{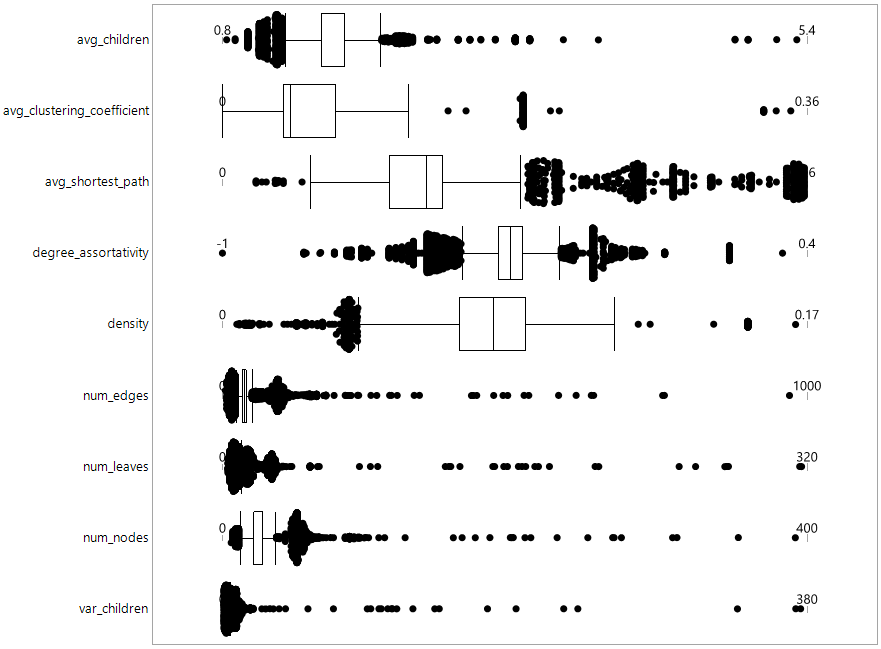}
    \caption{Box plot of the malicious PDFs. Features are on the y-axis.  }
    \label{fig:malPlot}
\end{figure}

Note that \autoref{tbl:beniQuan}, \autoref{tbl:beniStat}, \autoref{tbl:malQuan} and \autoref{tbl:malStat}
show a significant difference in the proposed feature set between benign and malicious PDFs. The statistical difference between benign and malicious PDFs is further visualized in the box plots ( \autoref{fig:BenPlot} and \autoref{fig:malPlot}), providing clear graphical representations of these differences. We excluded the median degree, depth, and median children from the box plots since  they are discrete values.

\section{Experimental Results}
\label{sec:results}

\subsection{PDF Malware Detection Results}
Having shown a significant difference in the proposed feature set between benign and malicious PDFs, we move on to demonstrate the feature set's utility in experimental evaluation. we use the Contagio dataset and the pdfrw library to extract tree structure paths for each file, and (as mentioned above) we were able to successfully extract path objects from the 7,396 benign PDF files and 10,814 malicious PDF files\cite{Xu2016AutomaticallyClassifiers}.  We use scikit-learn to evaluate the model's performance\cite{Pedregosa2012Scikit-learn:Python}. The confusion matrix we used is shown as \autoref{tbl:confmtx}
\begin{table}[!ht]
\centering
\caption{Confusion Matrix.}
\begin{tabular}{@{}lcr@{}}
\toprule
           & \multicolumn{1}{c}{Detected as Benign} & Detected as Malicious              \\ \midrule
Benign    & \multicolumn{1}{r}{True Positives (TP)} & False Negatives (FN)\\
Malicious    & \multicolumn{1}{r}{False Positives (FP)} & True Negatives (TN)\\ \bottomrule
\end{tabular}
\label{tbl:confmtx}
\end{table}

We apply 5-fold cross validation to 6 machine learning classifiers trained with the proposed 12 graph features. To show that our results are statistically significant, we report 95\% Confidence Interval (95\% CI) of the Precision, Recall in Table \ref{tbl:recall} and 95\% CI F1 score in Table \ref{tbl:f1}. The recall value indicates the ratio of samples have been correctly classified. We report the best recall value for the malicious class with Random Forest, indicating that 99.73\% to 99.77\% of the malicious samples are correctly detected, while the benign class's recall value is 0.9987 - 0.9991, which means that 99.87\% to 99.91\% of benign samples are correctly classified.

\begin{table}[!ht]
\centering
\caption{95\% CI Precision and Recall for each label.}
\addtolength{\tabcolsep}{-3pt}  
\begin{tabular}{@{}lcrcr@{}}
\toprule
                                      \\ 
         Classifier  & Label                      & \multicolumn{1}{c}{Precision} & Recall                       \\ \midrule
\multirow{2}{*}{XGBoost}    & \multicolumn{1}{r}{Malicious} & 0.9939 -	0.9977                   & \multicolumn{1}{r}{0.9958 - 0.9966}                    \\
   & \multicolumn{1}{r}{Benign} & 0.9982 - 0.9987                   & \multicolumn{1}{r}{0.9973 - 0.9991} 
                      \\ 
\multirow{2}{*}{Naive Bayes}    & \multicolumn{1}{r}{Malicious} & 0.8769 - 0.9564                    & \multicolumn{1}{r}{0.8000 - 0.9750}                \\
   & \multicolumn{1}{r}{Benign} & 0.8698 - 0.9683                    & \multicolumn{1}{r}{0.7850 - 0.9824}
                      \\
\multirow{2}{*}{Multi-layer Perceptron}    & \multicolumn{1}{r}{Malicious} & 0.9871 - 0.9911                    & \multicolumn{1}{r}{0.9811 - 0.9940}                 \\
   & \multicolumn{1}{r}{Benign} & 0.9946 - 0.9963                    & \multicolumn{1}{r}{0.9946 - 0.9963}
                      \\
\multirow{2}{*}{Decision Tree (J48)}    & \multicolumn{1}{r}{Malicious} & 0.9963 - 0.9973                    & \multicolumn{1}{r}{0.9946 - 0.9982}                  \\
   & \multicolumn{1}{r}{Benign} & 0.9972 - 0.9980                    & \multicolumn{1}{r}{0.9959 -	0.9986}
                      \\
\multirow{2}{*}{Random Forest}    & \multicolumn{1}{r}{Malicious} & 0.9966 - 0.9982                    & \multicolumn{1}{r}{0.9973 - 0.9977}                  \\
   & \multicolumn{1}{r}{Benign} & 0.9987 - 0.9991                    & \multicolumn{1}{r}{0.9987 - 0.9991} 
                      \\
\multirow{2}{*}{Simple Logistic}    & \multicolumn{1}{r}{Malicious} & 0.9739 - 0.9879                    & \multicolumn{1}{r}{0.9820 - 0.9824}                  \\
   & \multicolumn{1}{r}{Benign} & 0.9878 - 0.9885                    & \multicolumn{1}{r}{0.9831 -	0.9917}
                      \\
\bottomrule
\end{tabular}
\label{tbl:recall}
\end{table}

\begin{table}[!ht]
\centering
\caption{95\% CI F1 Score for each label.}
\begin{tabular}{@{}lcrcr@{}}
\toprule
                                      \\ 
         Classifier  & Label                      & \multicolumn{1}{c}{F1 Score}                        \\ \midrule
\multirow{2}{*}{XGBoost}    & \multicolumn{1}{r}{Malicious} & 0.9953 - 0.9968   \\
   & \multicolumn{1}{r}{Benign} & 0.9980 - 0.9986   \\ 
\multirow{2}{*}{Naive Bayes}    & \multicolumn{1}{r}{Malicious} & 0.8712 - 0.9234   \\
   & \multicolumn{1}{r}{Benign} & 0.8671 - 0.9227 \\
\multirow{2}{*}{Multi-layer Perceptron}    & \multicolumn{1}{r}{Malicious} & 0.9861 - 0.9906  \\
   & \multicolumn{1}{r}{Benign} & 0.9946 - 0.9963 \\
\multirow{2}{*}{Decision Tree (J48)}    & \multicolumn{1}{r}{Malicious} & 0.9959 - 0.9972 \\
   & \multicolumn{1}{r}{Benign} & 0.9970 - 0.9979  \\
\multirow{2}{*}{Random Forest}    & \multicolumn{1}{r}{Malicious} & 0.9970 - 0.9979  \\
   & \multicolumn{1}{r}{Benign} & 0.9987 - 0.9991  \\
\multirow{2}{*}{Simple Logistic}    & \multicolumn{1}{r}{Malicious} & 0.9781 - 0.9849  \\
   & \multicolumn{1}{r}{Benign} & 0.9854 - 0.9901  \\
\bottomrule
\end{tabular}
\label{tbl:f1}
\end{table}
We report the Accuracy, True Positive Rate (TPR), False Positive Rate (FPR), False Negative Rate (FNR) and True Negative Rate (TNR) in Table \ref{tbl:TPR}. The result indicates our proposed features have good overall performance. We report the best accuracy of 0.9975 when using Random Forest model.

   
   
  
  
 
  
\begin{table}[!ht]
\centering
\caption{Accuracy, True Positive Rate (TPR), False Positive Rate (FPR), False Negative Rate (FNR) and True Negative Rate (TNR) with the Proposed Features Set.}
\addtolength{\tabcolsep}{-3pt}  
\begin{tabular}{@{}lcrcrc@{}}
\toprule
                                      \\ 
         Classifier  & Accuracy                      & \multicolumn{1}{c}{TPR} & FPR                      & \multicolumn{1}{c}{FNR} & TNR\\ \midrule
XGBoost    & \multicolumn{1}{r}{0.9964} & 0.9980                    & \multicolumn{1}{r}{0.0042} & 0.0020 & 0.9958                    \\
   
Naive Bayes    & \multicolumn{1}{r}{0.9006} & 0.7850                    & \multicolumn{1}{r}{0.0268} & 0.2150 & 0.9732                   \\
   
Multi-layer Perceptron    & \multicolumn{1}{r}{0.9926} & 0.9919                    & \multicolumn{1}{r}{0.0088} & 0.0081&0.9912  \\
  
Decision Tree    & \multicolumn{1}{r}{0.9967} &0.9959                    & \multicolumn{1}{r}{0.0019} & 0.0041&0.9981                  \\
  
Random Forest    & \multicolumn{1}{r}{0.9975} & 0.9980                    & \multicolumn{1}{r}{0.0023} & 0.0020&0.9977                  \\
 
Simple Logistic    & \multicolumn{1}{r}{0.9822} & 0.9817                    & \multicolumn{1}{r}{0.0180} & 0.0183 & 0.9820                    \\
  
\bottomrule
\end{tabular}
\label{tbl:TPR}
\end{table}

\subsection{Comparison with Other Works}
For model performance comparison, we select models with reported results in the literature for direct comparison. 

\begin{table}[!ht]
\centering
\caption{Comparison with other models.}
\begin{tabular}{@{}lcrcr@{}}
\toprule

          Model & Precision                      & \multicolumn{1}{c}{Recall} & F1-Score                      & \multicolumn{1}{c}{Accuracy} \\ \midrule
Model\cite{Yerima2022MaliciousSet}    & \multicolumn{1}{r}{0.9970} & 0.9970                    & \multicolumn{1}{r}{0.9970} & 0.9965                     \\
Model\cite{AbuAl-Haija2022PDFTrees}    & \multicolumn{1}{r}{0.9880} & 0.9890                    & \multicolumn{1}{r}{0.9885} & 0.9884                     \\
Model\cite{Issakhani2022PDFLearning}    & \multicolumn{1}{r}{0.9888} & 0.9887                    & \multicolumn{1}{r}{0.9877} & 0.9869                    \\
Our work & \multicolumn{1}{r}{0.9973} & 0.9974                    & \multicolumn{1}{r}{0.9974} & 0.9975                      \\ \bottomrule
\end{tabular}
\label{tbl:MPR}
\end{table}


We report the best result obtained by applying the Random Forest model. \autoref{tbl:MPR} presents the comparison of the evaluation metrics to other work. The results show that our work beats other work while being significantly smaller than the feature set they use.

The main weaknesses of the feature set we propose is that it is vulnerable to some evasive attacks. Detectors that employ this feature set can be compromised through the insertion, deletion, or alteration of a subtree. In order to enhance the robustness of the detector, one potential strategy is to enrich the feature set and diversify the feature types employed.
We observe that the parser did not successfully parse all objects for some malware specimens.  The effectiveness of the our approach is influenced by the quality of parsed PDF objects. 

\section{Conclusion} \label{sec:conclusion}

In this work, we introduced a new features set for PDF malware detection that based on PDF tree structure. Our work aimed to address the need of finding a small features set without needing too much domain knowledge of the PDF file. Our work might serve as a baseline for the future investigation. We do not expect our work can replace the current used static and dynamic features now or in the future, but our work might inspire researchers with an alternative way to build a malware detection model. In the future task, we plan to explore the other features to improve the overall performance and enhance the robustness.


\bibliographystyle{ACM-Reference-Format}
\bibliography{references, local}

\end{document}